# Mapping Husserlian phenomenology onto active inference

Mahault Albarracin[1,2,*], Riddhi J. Pitliya[1,3,*], Maxwell J. D. Ramstead[1,4,†], and Jeffrey Yoshimi[5,†]

[1] VERSES Research Lab, Los Angeles, California, 90016, USA
[2] Université du Québec à Montréal, Montréal, Québec, Canada
[3] Department of Experimental Psychology, University of Oxford, Oxford, UK
[4] Wellcome Centre for Human Neuroimaging, University College London, London WC1N 3AR, UK
[5] University of California, Merced, 5200 Lake Rd, Merced, CA 95343, United States
* These authors contributed equally as first authors.
†These authors contributed equally as last authors.

**Abstract.** Phenomenology is the rigorous descriptive study of conscious experience. Recent attempts to formalize Husserlian phenomenology provide us with a mathematical model of perception as a function of prior knowledge and expectation. In this paper, we re-examine elements of Husserlian phenomenology through the lens of active inference. In doing so, we aim to advance the project of computational phenomenology, as recently outlined by proponents of active inference. We propose that key aspects of Husserl's descriptions of consciousness can be mapped onto aspects of the generative models associated with the active inference approach. We first briefly review active inference. We then discuss Husserl's phenomenology, with a focus on time consciousness. Finally, we present our mapping from Husserlian phenomenology to active inference.

## 1   Introduction

In recent years, there has been a resurgence of work attempting to formalize the structure and content of first-person conscious experience, leveraging mathematical and computational techniques to help model conscious experience [1]– [4]. One recently proposed version of this project, called "computational phenomenology", leverages the generative modeling techniques that were originally developed in computational neuroscience and theoretical neurobiology to formalize and model the structure and contents of conscious experience [5].

This paper aims to contribute to the project of computational phenomenology, by mapping core elements of the structure of conscious experience as described by Husserlian phenomenology to the constructs of the active inference framework, and in particular, to components of the generative models that underwrite that formulation. Computationally modelling conscious first-person experience using active inference would shed light on subjective individual experience and intersubjective experiences, which could be used to better understand factors constituting normal and abnormal behavior. We begin with a brief overview of active inference and generative modeling. We then review some of the core elements of Husserlian phenomenology, with a focus on time consciousness, drawing on the formalization of Husserl presented in [6]. We argue that we can use the generative models of active inference to represent these phenomenological structures. In so doing, we aim to advance the agenda for a computational phenomenology and take first steps towards worked examples of the method.

## 2    An overview of active inference

Given the intended audience of this paper, we will only briefly review active inference. In the broadest sense, active inference is a corollary of the free energy principle in Bayesian mechanics. Active inference is a process theory that can be used to model any physically separable, re-identifiable thing or particle, i.e., anything that persists as a coherent locus of states or paths, over some appreciable timescale. Active inference describes the dynamics (i.e., observable behavior) of things, so defined, as a path of least action, where the action is defined as time or path integral of an information theoretic quantity called self-information or, more simply, surprisal [7]–[11]. This quantity is also known as the negative log evidence in Bayesian inference. This means that the paths of least action maximize model evidence—a normative behavior sometimes referred to as self-evidencing [12]. In many practical applications of active inference, we do not consider the surprisal directly as it is often computationally intractable, since it requires averaging over a potentially infinite amount of states. Instead we consider an upper bound on surprisal called "variational free energy" [13]. This variational free energy measures the discrepancy between the observations or data that were expected, given a probabilistic (generative) model of how they were generated, and the data that was obtained. Intuitively, the idea is that any entity described by Bayesian mechanics maintains a model whose predictions tend to be confirmed over time (it minimizes the degree to which it is surprised). We will see that this kind of self-evidencing has a straightforward interpretation in Husserl's phenomenology.

In the narrower sense that will concern us more directly in this paper, the term "active inference" refers to a family of a mathematical models that we can use to simulate and model the behavior of cognitive agents [14], [15]. Active inference is usually implemented using partially observable Markov decision processes (POMDPs), or (equivalently) using Forney-style factor graphs [16]. In active inference, the action-perception loops via which agents engage with the salient features of their environmental niche, and with the other denizens of that niche, are cast as implementing approximate Bayesian (variational) inference (see [17] for a helpful introduction). Active inference thus comprises a set of formal tools, usually implemented in code, used to model the behavior of agents that interact with their environment, as a form of inference. The active inference toolkit allows us to model the epistemic and pragmatic imperatives of the behavior of agents: agents act to gather information about their environment and select those actions that bring them closer to characteristic states, which can be read as allostatic or homeostatic set points. In active inference, these set points—or attracting sets— are defined with respect to the kind of sensory data or outcomes that an agent expects to generate via action, given "the kind of thing that it is" [9].

Active inference is a situated or enactive kind of generative modeling, which considers not only how data are modeled—i.e., explained—but, crucially, how those data are gathered in the service of self-evidencing (see [5] for a review and discussion of its applications to phenomenology). Generative modeling underwrites many forms of mathematical modeling and scientific investigation [5], [18]. The general idea is straightforward. We have some data of interest, which we want to explain using statistical methods; i.e., we want to understand the *causes* of the data. So, we compute a number of alternative probabilistic models of the process that generated that data, and evaluate the evidence that the data provides for each model. In active inference, we assume that agents implement generative models, and update those models in light of sensory evidence. This modeling strategy assumes that agents can only access their environment by sampling it via sensory states. These generative models harness the beliefs of an agent about the "hidden" states of the external world, i.e., they encode what an agent knows about the process that generates its sensory data [19]. Agents are thus modeled as inferring what the primary causal pathways in the world are, and as navigating the opportunities for engagement that they are presented with by leveraging these inferences. Prior beliefs are updated continuously based on new data (i.e., new observations) via approximate Bayesian (variational) inference [20]. The current "content" of an agent's "experience" of "things" in the world is thus the set of states that are being inferred, on the basis of sensory data.

In active inference, *action* is modeled as a kind of self-fulfilling prophecy: agents predict what state they will be in upon acting, and then generate evidence for this prediction by actually acting in the environment [21]. The action itself is selected based on beliefs about possible courses of action, which are called "policies". Policies are thus beliefs about expected sequences of actions, which depend on an agent's beliefs about the current state of the world and the goals that it is trying to achieve (specified in terms of preferred observations). Different policies are, in some sense, variations of beliefs about expected future observations, contingent on possible courses of action. The value of a policy is determined by estimating a quantity is known as "expected free energy", which encodes how much each policy will minimize surprisal or, equivalently, maximize model evidence, with respect to preferred outcomes [22]. This rests upon the degree to which expected surprisal (i.e., uncertainty) can be resolved on the one hand, and the avoidance of surprising (i.e., aversive) outcomes on the other. We say that the optimal policy is the one that provides the most evidence for the generative model of the organism (or equivalently, that is expected to generate the least free energy). The selection of a policy is thereby driven by the expected free energy of that policy and agent's preferences, allowing an agent to conduct goal-directed behaviors.

To simulate an agent, we equip it with the states and parameters shown in Table 1, which can either be specified *a priori* by the experimenter, or learned based on real data [23]. In the POMDPs used in active inference, a distinction is made between observable data (denoted **o**), and hidden states (denoted **s**) [24]. The probability of some observation, given that some state obtains, is described by the likelihood matrix, denoted **A**; the entries of this matrix quantify the probability of observing some data, given that world is in some state. The parameter encoding the beliefs of the particle or thing about how states transition into each other over time is a matrix denoted as **B**, with each entry scoring the probability of transitioning to some state, given that the system was previously in some other state. A vector denoted **C** encodes preferences for each observation. Prior beliefs about base rates of occurrence of states are described by the **D** vector, with each entry scoring the prior probability of the associated state. Finally, baseline preferences for policy selection are described by the **E** vector. **C** is used to compute variational free energy (**F**) and expected free energy (**G**), which are used in perceptual inference and policy selection, respectively [25].

At any time-step, the current state is estimated by using "forward" and "backward" message passing. Forward messages are passed from nodes encoding beliefs about past states and observations to the node computing the current state; whereas backwards messages are passed from nodes encoding beliefs about future states and observations, contingent on policy selection, to the node computing the current state. To clarify, the agent does not "experience" the parameters of its generative model—encoded in its **A**, **B**, **C**, and **D** parameters. Rather, these parameters underwrite the message passing and belief updating; namely, the updating of prior beliefs into posterior beliefs in the face of new sensory evidence. As indicated, at any time step, the current "content" of an agent's "experience" of "things" in the world is implemented the set of states that are actively being inferred by the agent, on the basis of sensory data received.

## 3   An overview of Husserl's phenomenology

We now review Husserl's phenomenological description of time consciousness and intentionality, and how they constitute experienced object [1]. We use "phenomenology" in the technical sense that is commonplace in philosophy, to refer to a general descriptive methodology for the study of the structure and contents of the conscious, first-person experience of a subject or agent (or what might be called a "stable cognizer") [27]. We are concerned here with phenomenology as articulated by its founder, Edmund Husserl, who described it as an attempt to provide rigorous descriptions of the structure of first-person experience[2].

We are primarily concerned with Husserl's account of *time consciousness* [30]–[33]. For Husserl, consciousness evinces what one might call a kind of "temporal thickness", which is the ultimate condition of possibility for the perception of any object whatsoever[3]. Husserl's descriptive analyses suggest that the core structures of time consciousness, which enable what he calls the "constitution" of objects in consciousness (i.e., their disclosure to an experiencing subject) is threefold, comprising what he calls "primal impression", "retention," and "protention".

Primal impressions correspond to experience of the immediate present. Suppose a melody plays, or that you walk around an oak tree. The currently perceived note in the melody, or the current visual experience of the oak tree, correspond to primal impressions. In these cases, there is an additional structure that informs the primal impression: what Husserl calls hyletic data, or *hyle* (from the Greek word for matter, or stuff). The *hyle* correspond to our sense of the melody and the oak tree as being real occurrences in the world beyond us—a raw presence that we cannot alter by an act of will. However, we do not experience the *hyle* directly. They inform our primal impressions, but are not literal constituents of those impressions. The melody and the oak tree that we experience reflect both our own "top down" understandings of these things and their "bottom up" presence. Thus, our primal impression is a hylomorphic compound of raw presence and interpretation[4].

---

[1] In this paper we are mapping from one complex domain to another complex domain: active inference is a complex and growing area, as is Husserl scholarship [26]. Within Husserl scholarship, it is inevitable that we rely on existing interpretations, which are subject to scholarly dispute. This is a first sketch of the broad outlines of the mapping, that we aim to enrich in later work. For example, there are number of potential correlates of retention and protention in active inference discussed below, and further work is needed clearly delineating these.

[2] We will not be concerned here with the thorny issues that attend the naturalization of phenomenology [3]. See [5], [28], [29] for a review.

[3] In this, Husserl is aligned with other philosophers of his time, including James [34] and Bergson [35], [36]. A detailed historical analysis of the many sources of and precedents for Husserl's account of time consciousness is [37].

Primal impressions are "temporalized" in the flow of consciousness. More specifically, interpreted hyletic data are formatted into retentions and protentions. Retention is the "still living" preservation of the contents of a now-past primal impression in our present consciousness. In the case of the melody, one is still conscious of the notes that have just been struck, just as one hears the present tone. A protention corresponds to our sense of what will come next in the melody. Together they produce a temporal depth or thickness that is a condition of possibility for experiencing a melody, rather than a sequence of disconnected notes.

Husserlian retentions and protentions are not explicit representations. They are implicit, immediate components of the temporal thickness of experience, which can be contrasted with explicit representations of remembered past events (what Husserl calls "recollections") and explicitly anticipated future events; where an event in some set of (possibly nested) lived experiences. Remembering an important life event, or looking forward to some planned future events, are themselves mental acts that are experienced in ongoing processes with their own temporal depth (thus as, an explicit recollection unfolds, retentions and protentions associated with that recollected moment unfold as well) .

The experiences which unfold in time consciousness inform the way we understand the world to be—they "constitute" our sense of the world. In particular, protentions are tacit anticipations or expectations about what will happen in the next moment. When what actually happens next is consistent with our anticipations, we experience *fulfillment*. When what happens is inconsistent with our anticipations, we experience *frustration* or surprise (these are technical terms in Husserl; as with retention and protention, they do not imply an explicit or focused awareness). Thus, our experience of temporally extended objects consists in a flow of anticipations and fulfillment/frustration of those anticipations [39]. Our inner time-consciousness thus at core consists in a dynamic process that anticipates what will be experienced next, based on what has just been experienced.

Husserl suggests that over time retentions fade away and "sediment", informing our understanding of the world. Similarly for fulfilled or frustrated protentions. If the melody was much different than we thought it would be, the experiences of surprise would change our background understanding of the melody, leaving a trace, so that the next time we experience the same melody, our expectations have adapted to the change. In this way, we build up a kind of *model of reality* that is in the background of our experience, generating the anticipations and temporal depth of time consciousness (see [6], [40]). The fact that experience reflects all the sediments of past process of time consciousness means that our consciousness is laden with past retentions, which Husserl believes shape the way that we anticipate future primal impressions.

---

[4] The question of what exactly hyletic data are is a matter of controversy. We rely on a reading derived from Føllesdal [38], who says "In acts of perception our senses play a role, providing certain boundary conditions." They "limit" what what we can experience in a moment, without being directly experienced (they must be animated or interpreted by noetic form before they are experienced).

One analytic tool that Husserl introduces to study the structure of sedimented background knowledge is what he calls a "horizon" or "manifold". The idea with a horizon is to begin with some object given (i.e., constituted) in experience, and then to imagine different ways that an experience of that object could continue to be experienced. Each possible continuation of the current experience will produce a different profile of fulfillment and frustration. If we only focus on fulfilling continuations—that is, on further experience that would not surprise us—we get what is called a "trail set" in [6]. Trail sets can be used to formalize Husserl's notion of a horizon, i.e., what our implicit understanding of an object is, beyond what we immediately see. Standing before the oak tree, we have some expectations of how it would look, were we to move around it. Those expectations are open, they are "determinably indeterminate" leeways (*Spielräume*). These trails present the oak tree as having more or less branches on its back side, different coloration patterns, etc. However, they do not contain experiences of the back side of the oak tree that would surprise us, like one where a sign was nailed to it, or it was covered in spray paint. If we explore the oak tree, and one of those things is seen, then a protention will be frustrated, and that frustration will sediment in background knowledge, so that in the future, we will not be surprised: the trail set changes, we now see it as an oak tree with spray paint on its back. (This learning rule has been formalized using Bayesian statistics, making it easily amenable to active inference modeling; see [6]).

Our account so far has focused on perceptions of things or events, like seeing an oak tree or hearing a melody, but it can be extended to arbitrary mental processes, like hearing, imagining, planning, and also to more complex dynamic processes, like learning how to dance or learning mathematics. In each case, experiences unfold, and time consciousness operates, creating anticipations which are fulfilled or frustrated. The results of these processes of frustration and fulfillment are then sedimented into background knowledge. In this way, we maintain and update models of the external world, of how to dance, of mathematics, of history, of our own values and future plans, etc. These different types of knowledge generate different kinds of horizons associated with different kinds of trail sets: ways we expect things to be, ways we expect our body to move, ways we expect a conversation to unfold, what we expect ourselves to do relative to our values, etc.

## 4   Mapping Husserlian time consciousness onto generative models in active inference

In this section, we map aspects of the generative models that figure in active inference to aspects of Husserl's phenomenology. See Table 1 for a list of the states and parameters of generative models in active inference. As indicated, our project is situated within the broader framework of computational phenomenology[5].

We can associate aspects of a generative model, represented as a POMDP, to aspects of Husserlian phenomenology. In generative models, inferences about the current state of the world are informed by beliefs about what past states were experienced, and also by beliefs about what future states will be. Technically, the messages that are used to update current beliefs about hidden states come from factors that represent beliefs about states in the past, and also from factors that represent beliefs about states in the future.

We can begin by associating observations **o** with hyletic data, hidden states **s** with perceptual experiences, and the various parameters of the POMDP (e.g., the likelihood matrix **A** and state transition matrix **B**) with sedimented knowledge. Recall that in active inference modeling, outcomes are data that agents aim to explain (or alternatively, that we scientists are trying to explain, in generative modelling more broadly). Hidden states are inferred from this data, as their causes. During perception, hidden states of the generative model are used to generate predictions, which are compared against actual observations; and the parameters of the model are updated in a Bayes optimal manner, such that these predictions get better over time, leading to reductions in variational free energy. This maps directly on to the Husserlian apparatus. Our immediate perceptual experiences (correlated with **s**) are based on a mixture of relevant background knowledge (correlated with **A**, **B**, and so on) and hyletic data (correlated with **o**). The hyletic data are not literal constituents of experiences, just as **o** is not a literal constituent of **s**. Rather, the hyletic data impose boundary conditions or limits on what we can experience—they correspond to a sense of the presence of the world—but they are not experienced directly. Sensory experiences arise from the *interplay* of hyletic data and background knowledge (or "noetic form") in Husserl. In a similar way, **o** constrains or limits what hidden state **s** will be inferred, given **A** and **B**, but is not contained in **s**. Hidden states are updated as a function of observed sensory states **o** and beliefs encoded in a likelihood and transition matrices **A** and **B**, but the hidden states do not directly contain those observations.

There are several ways to capture retention and protention in a generative model. One way is to focus on the process of inferring hidden states from observed

---

[5] Technically, computational phenomenology is a version of generative modeling that is agnostic about whether the models at play are real descriptions of the actual processes at play in agents, or whether these models are merely useful heuristics to model first-person experience. See [5] for a discussion. Of note, the work presented in this paper dovetails nicely with realist approaches the implementation of generative models by agents,; see integrated world modelling theory as proposed in [2]

data, by comparing a prediction about what will be observed with what is actually observed, and updating beliefs using an error signal (as in predictive coding implementations of active inference, [41]). Such an approach involves direct correlates of protention (a prediction signal), fulfillment or frustration (the error signal), and beliefs update on that basis. Thus, even in the immediate or "static" perception of an oak tree [15] the process of state estimation involves correlates of protention, fulfillment and frustration. A second approach is to focus on the state transition matrices **B**, which encode state transition probabilities, and which thus underwrite "dynamic perception" [15], that is, beliefs about how objects change over time. These matrices are used to estimate what will occur next in a song as we listen to it, or how the oak tree will sway under the influence of the wind. These state estimates themselves rely on what occurred just previously (see, e.g., Figure 4 in [42]). So here again we have direct correlates of retention and protention in the active inference framework, this time in the operation of the **B** matrices. There may also be links between retention and working memory, especially when it is understood (in an active inference framework) in terms of evidence accumulation in a temporally structured hierarchy [43][6].

The experience of fulfillment and frustration can be modelled as a process of Bayesian belief updating [6]. In line with this analysis, we suggest that we can quantify fulfillment/frustration in terms of the variational/expected free energy that is generated by subsequent sensory experience; where the free energies quantify the degree to which current experiences conflict with protended experience (i.e., the degree of fulfillment or frustration).

Active inference, like Husserlian phenomenology, is ultimately about action in a lived world. As described above, a policy in active inference is a set of beliefs about possible courses of action; and action itself is modelled as a kind of self-fulfilling prophecy. This basic structure can be extended to include counterfactual richness, which can be associated with the trails of fulfilling experiences in a horizon. In so-called "sophisticated" treatments of active inference, agents select which action to pursue by engaging in a deep tree search, unfolding possible sensory consequences of available actions recursively, and evaluating each branch in terms of the free energy expected along that branch [46]. The search process is efficient; only those paths with high posterior probability are evaluated, but the search is defined over a larger set of possible paths. The optimal policy is the one that maximizes preferred outcomes (relative to **C**, which encodes prior preferences for data) and maximizes model evidence (or minimizing surprise). These counterfactual policy deep trees of sophisticated active inference can be mapped to Husserlian structures, including a set of values encoded in background knowledge (a correlate of **C**), and other features of background knowledge (e.g., our knowledge of state transitions, encoded in **B**), which can be used to generate a trail set: a set of expected perceptions that is consistent with our beliefs, goals and desires.

---

[6] A fuller discussion would also involve a comparison with existing discussions of the naturalization of time-consciousness, such as [44] and [45].

We can thus describe a mode of analyzing POMDPs that maps onto the method of horizon analysis in Husserl. Focusing on the case of perception, imagine all possible sequences of data **o**, given some assumed hidden state **s**, and a policy or sequence of actions. Some of the data generated will confirm the knowledge implicit in the model parameters (i.e., provide evidence for it, as quantified by free energy); others will disconfirm it (i.e., will generate large amounts of free energy). The sequences of observations that confirm the beliefs of the agent about the current state of its world correspond to a set of possible continuations from the current observation that are not surprising, i.e., that lead to little variational free energy. Such continuations are captured in the parameters of the generative model (e.g., the **A** and **B** matrices). These active inference trail-sets map directly on to the Husserlian trail sets. The latter can be thought of as an alternative, and perhaps more intuitive, way of understanding the information implicit in the matrices. In the one case, we have a method of "probing" the expectations implicit in the parameters of the generative model; in the other case, we have a method of probing the expectations implicit in an experiential horizon.

The representational analogs of retention and protention (recollection and explicit prediction) can be formalized by appealing to (possibly hierarchical) state estimation. Indeed, focally recollected and anticipated events constitute (past and future) states of the world (or indeed, of the self) that need to be represented explicitly. To begin to formalize this, one can point to the explicit distinction, in active inference, between the **A**, **B**, **C**, and **D** parameters, which contribute to current state estimation, and the states which are actually being inferred in the present, to account for the distinction. This richness of this account can be increased by appealing to nested hierarchies of generative models. Explicit recollection of memories and anticipation would then correspond to state factors higher up in the hierarchy, which bin or coarse-grain observations at subordinate layers.

In both active inference and phenomenology, the analysis of perception is just a convenient starting place. All the mappings developed above can be applied to other features of cognition and experience: auditory and tactile experience, multi-modal experience, cognition, language, skilled behavior, planning, affect, intersubjectivity, etc., each associated with its own state estimations, learned likelihood matrices, retentions and protentions, trail sets, and so forth.

## 5   Conclusion

This paper has drawn parallels between Husserlian phenomenology and the active inference framework. We proposed to formalize some core elements of Husserlian phenomenology via active inference. Our aim in so doing was to advance the project of computational phenomenology. We proposed Husserl's descriptions of primal impression, hyletic data, retention, protention, fulfillment, frustration, trail set and horizon, recollection and explicit anticipations can be mapped onto aspects of the generative models of active inference.

Husserlian phenomenology is fertile ground for formalization. Formalizing phenomenology allows us to leverage it in order to better understand and model human experience, and to make testable empirical predictions. Concurrently, active inference has been used to model many aspects of cognition, but its use to explain qualitative and subjective experience is still in the very early stages . Moving towards computational phenomenology through a connection between Husserlian phenomenology and active inference may allow us to bridge the gaps to fundamental questions such as the explanatory gap and positionality, and extend further into sociological issues of intersectionality, which make fundamental reference to first-person experience.

### Acknowledgments

The authors thank Philippe Blouin, Laurence Kirmayer, Magnus Koudahl, Antoine Lutz, Jonas Mago, Jelena Rosic, Anil Seth, Lars Sandved Smith, Dalton Sakthivadivel, and the members of the VERSES Research Lab for useful discussions that shaped the contents of the paper. Special thanks are due to Juan Diego Bogotá, Zak Djebbara, and Karl Friston.

# 7   Tables

**Table 1.** Parameters used in the general model under the active inference framework. We explain the generative model symbols that refer to different matrices and elements which are connected through them.

| | |
|---|---|
| **o** | Observations or sensory states of an agent |
| **s** | Hidden or external states |
| **A** | Likelihood matrix that captures beliefs about the mapping from observations to their causes (hidden) |
| **B** | Transition matrix that captures beliefs about the mapping between states at one time step to states at the next time step |
| **C** | Prior preference matrix that captures the preferred observations for the agent, which will drive their actions |
| **D** | Priors that capture beliefs about base rates of occurrence of the hidden states |
| **E** | Prior preferences for policies in the absence of data |
| **F** | Variational free energy |
| **G** | Expected free energy |
| $\pi$ | Policy matrix that captures the policies available to an agent |